Session 5: Optimal Experimental Design

**Optimal Sampling & Reconstruction: Theory and Applications**

**Justin Haldar <jhaldar@usc.edu>**
**Signal and Image Processing Institute**
**Ming Hsieh Department of Electrical and Computer Engineering**
**University of Southern California**

**Problem Summary**

The optimization of MRI data sampling and image reconstruction methods has been a priority for the MRI community since the very early days of the field. Designing an "optimal" method requires the definition of an optimality metric (i.e., a quantitative evaluation of the "goodness" of different competing approaches that allows an objective comparison between them). However, a key challenge is that there are many different possible ways of quantitatively evaluating the "goodness" of a data sampling scheme or a reconstruction result, and there are no acquisition or reconstruction methods that are known to be universally optimal with respect to all of these possible metrics simultaneously. Thus, optimization of MRI methods requires a subjective choice about what aspects of quality matter most in the context of a given MRI experiment, and subsequently the subjective choice of an optimality metric that hopefully does a reasonable job of quantifying those aspects of quality. Once these choices are made, the optimization problem becomes well-defined, and it remains to choose an algorithm that can identify data sampling or image reconstruction methods that are optimal with respect to the chosen metric. All of these choices are generally nontrivial.

In this presentation, we will discuss optimal sampling and reconstruction designs from multiple different perspectives, including ideas from information and estimation theory and various practical perspectives.

**Description**

It is common to assume that the ultimate goal of an optimal imaging experiment should be to acquire data and reconstruct images that are as informative as possible, subject to practical experimental constraints. However, while this may seem reasonable at first, it quickly becomes apparent that the statement is not very meaningful unless the concept of "information" is defined in a precise way that can be quantified and optimized. One such technical definition of information was introduced in 1948 by Claude Shannon in a landmark paper that inaugurated the field of Information Theory[1].

Shannon's concept of information is intimately connected to uncertainty. Specifically, a data measurement will contain information if it reduces our uncertainty about the image (in an appropriate probabilistic sense), and an information-theoretically optimal experiment would take measurements that are ultimately expected to result in the smallest possible amount of uncertainty.

For our purposes, it is important to note that the concept of uncertainty depends rather critically on what is already known. For example, the answer to the question "did it rain today?" would yield more information if we're asking about a specific location on Earth (where the weather can vary substantially from day to day) than if we're asking about a specific location on Pluto (where we could predict the answer with almost 100% certainty based on prior knowledge). Similarly, the question "did he survive?" would provide much more information if we were asking about a person who had a heart attack compared to a person who had a paper cut. Thus, in the context of experiment design, the most informative data samples will be those samples that are the hardest to predict based on what we already know.

At the same time, not all uncertainty is equally important for us to resolve – for example, in a head MRI scan that is designed to provide information about the brain, we may not care if there is substantial uncertainty about the portion of the image that is associated with the reconstruction of the eye. Parameters that we don't care about are often classified as "nuisance parameters." Nuisance parameters are important to consider because they impact our ability to estimate parameters of interest -- however, gaining information (i.e., reducing uncertainty) about such nuisance parameters does not need to be a priority for optimal design if that information is not relevant to our ultimate imaging goals.

In order to apply these concepts to the optimization of MRI, we therefore need to formalize what we do and do not already know, while also developing a clear understanding of what information is important to know versus the information that may not be relevant.  Unfortunately, such formalization is very hard to achieve, because our prior knowledge and priorities can vary dramatically from one imaging context to the next, and also because different people often have different views about what might be important.  For example, radiologists, statisticians, physicists, and engineers will often not express concern or uncertainty about the same things when presented with a set of MRI data.

These complexities make it hard to define optimization formulations that truly capture our uncertainties and our imaging goals, and it is common for researchers to optimize sampling and reconstruction methods in abstracted/simplified scenarios without attempting to incorporate all of the relevant knowledge that we may have.  For the purposes of this presentation, we group optimal sampling and reconstruction design methods into three groups: "model-free," data-driven, and model-based design methods.

We define "model-free" design methods as methods that make minimal assumptions about the characteristics of an image, and instead attempt to optimize theoretical image quality characteristics that transcend any specific dataset or application (e.g., the shape of the point-spread function (PSF) and the signal-to-noise ratio (SNR)).   Frequently, such methods rely on continuous (infinite-dimensional) representations of the image and use simple linear reconstruction techniques that are easy to characterize theoretically. (Note that these "model-free" approaches are still actually based on some modeling assumptions, despite the terminology we are using.  But they are based on much less restrictive assumptions compared to the model-based methods we will discuss in the sequel).  Examples include sampling-design and reconstruction-design methods that aim to achieve an optimal trade-off between spatial resolution (e.g., as measured through the PSF) and noise variance, under constraints on experiment duration[2-9].  Interestingly, although the issue of the optimal trade-off between spatial resolution and SNR had been "solved" in classical MRI literature[10,11], modern theoretical analyses are overturning some of the conventional wisdom in the field[8].  For instance, it may not be common knowledge, but it can be shown theoretically that for simple linear reconstruction methods, if the optimization objective is to optimize SNR (assuming a fixed experiment duration and a fixed spatial resolution), then it is often more efficient to apply spatial smoothing to noisy data acquired at higher-than-necessary resolution than it is to acquire multiple averages of data acquired with exactly the desired resolution[8,12-14].  This kind of observation can be very important in the context of SNR-starved MRI applications.

Conversely, data-driven design methods rely on application-specifc prior information.  Since it is difficult to formalize the likelihood of encountering MRI images with specific characteristics, these approaches generally use a database of existing MR images in order to empirically sample from the distribution of typical images.   For example, in the modern model-based MRI reconstruction literature, sampling patterns and reconstruction hyperparameters are often optimized by computing empirical errors with respect to a database of existing images.  And recently, data-driven approaches to image reconstruction and data-sampling design are becoming increasingly popular due to recent the emergence of machine-learning and deep-learning methods.  For these methods, the underlying assumption is that prior information about what is known or unknown about an image can be extracted automatically from previous examples, and such prior information can then be used to design better sampling and reconstruction methods. (As a historical note, it is worth observing that methods that use a database of previous images to design optimal sampling and reconstruction methods have existed in the field for a long time[15], although the early contributions aren't very well-known today). However, a potential limitation is that the prior information that is learned from the data is frequently opaque to the user. As a result, data-driven approaches may potentially suffer from unexpected biases and blindspots, which may lead to problematic failure cases.

The last category of methods uses explicit modeling assumptions, which allows optimal experiment design and image reconstruction to be formulated using statistical tools.  For example, it can be common to adopt a constrained model that reflects the prior information that most images are likely to be support-limited, have smooth phase, do not have much information at high-frequency k-space locations, are sparse in a transform domain, are acquired with an array of receiver coils that possess complementary information, etc.[16-19]. Given such an image model together with a data acquisition/noise model, there are well established techniques in estimation theory/statistics that can be used to design both optimal estimation/reconstruction strategies[20] as well as optimal data sampling strategies[21].

On the image reconstruction front, there are different common ways of defining statistical optimality that lead to different formulations of the image estimation problem[20]. Common choices include maximum-likelihood, penalized maximum-likelihood, and Bayesian estimation methods, and all of these are commonly used in MRI. However, the notions of optimality that are currently used with such approaches are often generic and do not account for the application-specific goals of the imaging experiment in question, meaning that there is room to refine the notions of optimality that we consider.

On the data sampling front, there are also different ways of defining statistical optimality, and a variety of different algorithms that can be used to identify optimal sampling schemes. A key theoretical concept for experiment design is the Cramér-Rao Bound (CRB), which provides an objective mechanism to measure the uncertainty associated with any given experiment design[20], and which is intimately connected to the well-known g-factor metric from the parallel MRI literature[22,23]. Sampling design methods that seek to optimize the CRB (or g-factor) have been applied with great success for both k-space sampling design[24-31] and more general sampling design for quantitative MRI in a number of different contexts[32-35]. However, as before, the current methods frequently do not account for the specific goals of the imaging experiment in question, meaning that there is also considerable room for improvement.

A fundamental issue for optimal design is the choice of an appropriate image quality metric. Commonly-used metrics include the root-mean-squared error (RMSE), peak signal-to-noise ratio (PSNR), and structural similarity (SSIM), which each have their own limitations[36]. However, alternative error metrics have recently been introduced that can provide more insight into different aspects of image quality[37] that offer potential improvements. In addition, application-specific task-based measures of image quality have been explored for imaging modalities outside of MRI[38], but there has been more limited translation of such concepts into MRI. Exploring such directions for MRI has the potential to be very fruitful.

Finally, we should mention that from an information theoretic point of view, the process of image reconstruction will never result in a gain in information, it can only cause a loss of information. This is a consequence of a theoretical result known as the "data processing inequality." This suggests that the process of image reconstruction is inherently suboptimal from an information theoretic point of view. Of course, image reconstruction is still important for human interpretation of imaging data, which is an important part of the imaging pipeline (at least at this point in history!).

**Conclusions**
The design of optimal sampling and reconstruction methods is an important problem that has been studied for decades. Although substantial advances have been made, a major hurdle for this research area is that it is difficult to define appropriate notions of optimality that embed both the prior knowledge that we have about a given MRI scan as well as the information that is most important to learn from the scan. These are difficult issues, and addressing them likely requires a continuing dialogue within the MRI community to clearly define exactly what our ultimate imaging objectives should be, and how to represent these objectives in a way that is amenable to mathematical optimization.